\documentclass[preprint,12pt,3p]{elsarticle}
\usepackage[utf8]{inputenc} 
\usepackage[T1]{fontenc}    
\usepackage{hyperref}       
\usepackage{url}            
\usepackage{booktabs}       
\usepackage{amsfonts}       
\usepackage{nicefrac}       
\usepackage{microtype}      
\usepackage{lipsum}
\usepackage{amsmath,amssymb,graphicx}
\usepackage{float}
\usepackage{siunitx}
\usepackage{array}
\usepackage{lmodern}
\usepackage{amssymb}
\usepackage{xcolor}
\usepackage{multicol}
\usepackage{multirow}
\usepackage{orcidlink}
\usepackage{amsmath,amssymb,exscale}
\usepackage{amsmath}
\usepackage{amsmath,amssymb,exscale}
\usepackage[final]{pdfpages}
\usepackage[bottom]{footmisc}
\journal{Elsevier}

\begin{document}
\begin{frontmatter}
\title{First-principles calculations of the electronic and optical properties of
penta-graphene monolayer: study of many-body effects}

\author[Damghan_University]{Babak Minaie\corref{author}\,\orcidlink{}}
\author[Damghan_University]{Seyed A. Ketabi\corref{author}\,\orcidlink{0000-0001-6629-0918}}
\cortext[author]{Corresponding authors: babminaie@std.du.ac.ir (Babak Minaie) and saketabi@du.ac.ir  (Seyed Ahmad Ketabi) and josemoreiradesousa$@$ifpi.edu.br (José M. De Sousa) }

\author[IFPI]{José M. De Sousa\corref{author}\,\orcidlink{0000-0002-3941-2382}}
\affiliation[Damghan_University]{organization={School of Physics},
            addressline={Damghan University}, 
            city={Damghan},
        country={Iran}}

\affiliation[IFPI]{organization={Instituto Federal de Educa\c c\~ao, Ci\^encia e Tecnologia do Piau\'i -- IFPI},
            addressline={Primavera}, 
            city={São Raimundo Nonato},
            postcode={64770-000}, 
            state={Piauí},
            country={Brazil.}}
                    
\begin{abstract}
In the present work, first-principles calculations based on the density functional theory (DFT), GW approximation and Bethe–Salpeter equation (BSE) are performed to study the electronic and optical properties of penta-graphene (PG) monolayer. The results indicated that PG is a semiconductor with an indirect band gap of approximately $2.32$ eV at the DFT-GGA level. We found that the utilization of the $GW$ approximation based on many-body perturbation theory led to an increase in the band gap, resulting in a quasi-direct gap of $5.35$
eV. Additionally, we employed the $G_{0}W_{0}-RPA$ and $G_{0}W_{0}-BSE$ approximations to calculate
the optical spectra in the absence and in the presence of electron-hole interaction,
respectively. The results demonstrated that the inclusion of electron-hole interaction caused a red-shift of the absorption spectrum towards lower energies compared to the spectrum obtained from the $G_{0}W_{0}-RPA$ approximation. With the electron-hole interaction, it is found that the optical absorption spectra are dominated by the first bound exciton with a significant binding energy $3.07$ eV. The study concluded that the PG monolayer, with a wider band gap and enhanced excitonic effects, holds promise as a suitable candidate for the design and
fabrication of optoelectronic components. 
\end{abstract}

\end{frontmatter}

\section{1. Introduction}
Carbon atom is capable of forming various allotropes at different dimensions due to its unique electron configuration. Graphite, diamond, fullerenes  \cite{dresselhaus1993fullerenes}, carbon nanotubes \cite{qin2000smallest}, graphene, and graphyne  \cite{li2014graphdiyne} are among the carbon allotropes. Each of these allotropes possesses unique and distinct properties compared to the others. The successful synthesis of graphene \cite{novoselov2004electric} has led to a great interest in the investigation of graphene-based nanomaterials among many researchers. Graphene, due to its two-dimensional structure and extremely low
thickness (approximately the diameter of a carbon atom), as well as its unique properties such
as high carrier mobility, thermal conductivity, and mechanical stability, holds great potential
for various industrial applications \cite{geim2007rise,geim2009graphene,rao2009graphene,zhu2010graphene,du2008approaching,balandin2008superior,grantab2010anomalous}.

However, despite the mentioned features, graphene is a zero-bandgap semiconductor, which significantly hinders its direct application in optoelectronic-based devices. It is notable that the structure-property relationship is an important fundamental research issue in the field of 2D materials. Therefore, extensive
theoretical and experimental studies have been conducted to modify the bandgap of graphene \cite{sahu2017band,silvestrelli2018bandgap,park2015band,khoa2019opening,ke2019large}. Various methods have been proposed to induce a bandgap in the graphene structure,
such as applying an electric field, strain, chemical functionalization and doping processes.
Another approach to modify the structural properties is the logical manipulation of the blocks
that form the desired structure, which can lead to a new class of materials with improved
characteristics. As a result, significant efforts have been devoted to the search for suitable
geometrical blocks beyond the conventional hexagonal ones. The design and synthesis of
pentagonal materials heavily rely on flexibility in bonding and hybridization of elements.
Carbon has emerged as the researchers first choice for this purpose due to its ability to form
single, double, and triple bonds and exhibit various hybridizations, including $sp$, $sp^{2}$ and $sp^{3}$. Furthermore, the successful isolation of $C_{20}$ fullerene, which consists solely of $12$ pentagons \cite{prinzbach2000gas}, has expanded the availability of other low-dimensional pentagonal materials.

Based on extensive computational simulations, in $2015$, penta-graphene (PG), a 2D carbon allotrope composed exclusively of pentagonal blocks, was proposed \cite{zhang2015penta} and subsequently, extensive studies on PG and its derived structures have been conducted  \cite{tien2020influence,yu2015comparative,li2016tuning}. Some previous studies have investigated the thermal transport properties \cite{xu2015thermal}, optical properties \cite{wang2016electronic}, and mechanical properties \cite{sun2016mechanical} of PG, as well as its electronic properties using approaches such as tight-binding model and DFT-GW methods \cite{stauber2016tight}. In a study conducted by Li et al. \cite{li2016tuning}, utilizing the VASP computational package and employing the generalized gradient approximation $(GGA)$ with the inclusion of Perdew-Burke-Ernzerhof $(PBE)$ and the Heyd-Scuseria-Ernzerhof $(HSE06)$ functionals, it was demonstrated that chemical functionalization, specifically hydrogenation and fluorination, can effectively modify the mechanical and
electronic properties of PG. Their studies demonstrate that chemical functionalization
increases the band gap of PG and, in particular, brings its semiconductor nature closer to insulators. In a similar work, Jia et al. \cite{jia2018piezoelectric} using first-principles calculations based on DFT,
investigated the electronic, mechanical, and piezoelectric properties of hydrogen- and fluorine-functionalized PG sheets. Their calculations demonstrate that the band gap of the PGsheet increases from $3.25$ eV to $4.86$ eV, $4.6$ eV and $4.42$ eV upon hydrogenation, hydro-
fluorination, and fluorination of the structure, respectively. In addition to chemical functionalization, the use of different approximations and functionals
for calculating the properties of PG can lead to different results. Zhi Gen et al. \cite{yu2015comparative} investigated the electronic properties of monolayer PG and compared the results with multilayer PG using the HSE06 and $optB88-vdW$ functionals. Their results showed that the calculated band gap using the $HSE06$ functional for PG with different layer numbers are all
direct, and increasing the number of layers did not change the gap from direct to indirect. Einolazadeh et al. \cite{einollahzadeh2016computing} studied the electronic properties of PG sheet in DFT-LDA level and
also GW-level.

Their calculations demonstrated that the inclusion of electron-electron
interactions through the $G0W0$ approximation leads to an increased band gap. Unlike graphene, which has been extensively studied to discover its various properties, and it can be
said that research related to it is almost saturated, the studying on PG structures are still at an early stage. Furthermore, most previous studies have focused on the obtaining the ground state properties of this structure, and the investigation of excited states and the properties of
the structure under many-body interactions, especially electron-hole interactions, has been
overlooked. Therefore, in this work, we intend to investigate the electronic and optical properties of the PG monolayer by combining DFT with the GW approximation and solving the Bethe-Salpeter equation. To achieve this, we will first calculate the ground state properties of the structure using density functional theory and then by considering the screened Coulomb electron-electron interaction through the GW approximation and solving
the Dyson equation, we will obtain the quasi-particle energies. Finally, by taking into account the electron-hole interaction and solving the Bethe-Salpeter equation, we will examine the effects of excitons on the band structure and optical spectra of the PG monolayer. The
description of the computational schemes is introduced in Section  \ref{Methodology}. The results and
discussion are presented in Section  \ref{Results}, followed by conclusions in Section  \ref{conclusion}.

\section{Computational Methodology}
\label{Methodology}

In this section, we present the computational methods based on which we performed our calculations to investigate the electronic and optical properties of PG. In the first step, for structural optimization and electronic-band structure calculations of PG monolayer, first principle calculations were performed in the framework of DFT as implemented in the QUANTUM-ESPRESSO computational package. The PBE \cite{perdew1996generalized,perdew1998erratum} functional of generalized gradient approximation (GGA) \cite{perdew1998perdew} was adopted to describe electron exchange-correlation.

Furthermore, the Martin-Taylor pseudopotential \cite{troullier1991efficient} which is of norm-conserving type was employed to investigate the electron-ion interactions. To ensure accurate and minimally
error-prone calculations, an optimal number of k-points of $14 \times 14 \times 1$ were chosen in the primitive unit cell of the PG monolayer. The convergence threshold for the total energy was
set to around $10^{-4}$ eV/atom, and the force convergence threshold was set to around $10^{-5}$ eV/$\AA$. A suitable energy cutoff of $50$ Ry was considered for this structure. According to the value of the bond lengths of $C_{1}-C_{1}$ ($1.550$ \AA) and $C_{2}-C_{2}$ ($1.329$ \AA), a vacuum distance of $15$ \AA in the
perpendicular direction is considered to avoid interactions with adjacent periodic structures. It
should be noted that DFT is only capable of calculating the ground state properties of the
system and has limitations in determining excited states. However, to calculate the band
structure and obtain the band gap of a system, we need at least one electron in the conduction
band, which itself is an excited state. Therefore, we need to go beyond density functional
theory. The GW approximation in the framework of many-body perturbation theory is based
on the Green's function formalism and quasi-particle concept  \cite{aulbur2000quasiparticle}.

The results obtained from the GW approximation for calculating the band gap of a many-body system show good
agreement with available experimental results [33] \cite{korbel2014benchmark}. Therefore, in the next steps, we also use
the GW approximation to calculate and determine the electronic properties of excited states. In order to study the quasi-particle energies within the many-body perturbation theory, the
exchange-correlation potential employed by DFT must be replaced by the self-energy
operator calculated within the GW approximation \cite{hedin1971explicit} on the basis of the generalized
Plasmon-pole Model \cite{godby1989metal}. The self-energy $\sum$ has been calculated as the product of one-electron Green's function $G_{0}$ and the screened coulomb potential $W_{0}$ , i.e., $\sum = i G_{0}W_{0}$ . To
obtain the optical properties of a structure, the behavior of various optical quantities as a
function of incident photon energy needs to be studied. Calculating the dielectric function,
which is a function consisting of both real and imaginary components, serves as a starting
point for describing the optical properties of a structure. Therefore, we intend to calculate the
dielectric function for PG monolayer using the following two approaches:

\begin{itemize}
    \item The random phase approximation based on the quasi-particle energies, where only
screened Coulomb interactions between electrons are considered ($G_{0}W_{0}-RPA$).\\

\item Solving the Bethe-Salpeter equation (BSE) based on the two-particle Green's
function, where in addition to electron-electron interactions, electron-hole interactions
are also taken into account ($G_{0}W_{0}-BSE$).
\end{itemize}
To solve the Bethe-Salpeter equation, we use the following relation \cite{onida2002electronic,rohlfing2000electron},

\begin{eqnarray}
(E_{ck}^{qp} - E_{vk}^{qp})A_{vck}^{s} + \sum_{k' v' c'} \left \langle v c k | K^{eh} | v' c' k' \right \rangle A_{v' c' k'}^{s} = \Omega^{s} A_{v c k}^{s},
\end{eqnarray}
where $E_{ck}^{qp}$ and $E_{vk}^{qp}$ denote the quasi-particle energies of conduction and bands at the specific wave vector, respectively. $K^{eh}$ is the eletron-hole screened interaction kernel. $A_{v c k}^{s}$ and $\Omega^{s}$ are the exciton eigenvalue, respectively.  $| v c k >$ and $| v' c' k' >$ denote the quasi-electron and the quasi-hole states, respectively. The calculations performed
using the YAMBO \cite{marini2009yambo} computational code.

\section{Results and discussion} 
\label{Results}

This Section presents the results of the calculations based on the formalism described in Section \ref{Methodology}. To continue, in the following subsections, we study the dynamical stability, the electronic and optical properties of the PG monolayer, such as phonon dispersion, electronic band structure, dielectric function and the other optical parameters.

\subsection{Phonon Dispersion and Dynamical Stability}

\begin{figure}[htb!]
\begin{center}
\includegraphics[width=0.90\linewidth]{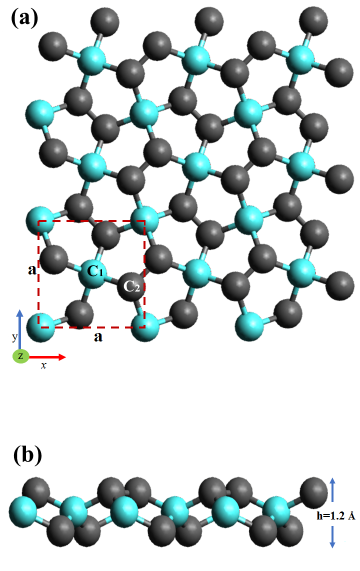}
\caption{{\textit{(Colour online) The plot shows schematically penta-graphene (PG) monolayer structure. (a) top
view and (b) side view. The square shown with dashed red lines is the primitive unit cell of PG.}}}
\label{FIG:01}
\end{center}
\end{figure}

Figs. \ref{FIG:01}(a) and \ref{FIG:01}(b) show schematically the top view and side view of the buckled PG
monolayer structure, respectively. After optimizing the structure using the QUANTUM-
ESPRESSO computational code, the lattice constant was obtained as $a = 3.64$ \AA, which is in perfect agreement with other results \cite{zhang2015penta}. The unit cell of the PG monolayer structure
consists of $6$ carbon atoms with $sp^{2}$ and $sp^{3}$ hybridizations is shown in Fig. 1a, where $C_{1}$
atoms have $sp^{3}$ hybridization and $C_{2}$ atoms have $sp^{2}$ hybridization.

\begin{figure}[htb!]
\begin{center}
\includegraphics[width=0.90\linewidth]{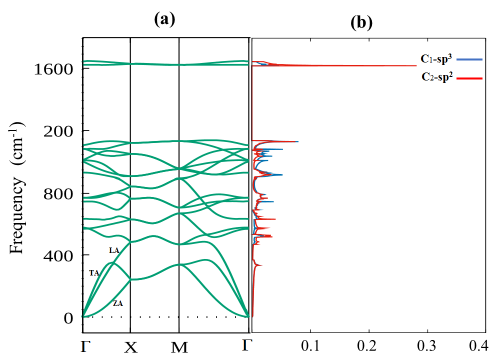}
\caption{{\textit{(Colour online) Panels (a) and (b) show phonon spectra and phonon density of states of the PG
monolayer, respectively.}}}
\label{FIG:02}
\end{center}
\end{figure}

The bond length between
$C_{1}-C_{1}$ atoms and $C_{2}-C_{2}$ atoms were obtained $1.550$ \AA~ and $1.329$ \AA, respectively. The
thickness of the buckled structure or the buckling height is calculated to be $h = 1.2$ \AA. Such
buckling can also be seen in other mono-atomic monolayer crystals such as silicene,
germanene, stanene and black phosphorus. To determine the dynamical stability of PG monolayer structure, the phonon spectrum of the
structure needs to be calculated. Panel (a) in Fig. \ref{FIG:02} illustrates the phonon dispersion curves
and the corresponding phonon density of states for the PG monolayer along the high-
symmetry points of the first Brillouin zone. Based on the phonon spectrum, it can be inferred
that the structure is stable since there are no negative phonon frequencies, indicating the
absence of imaginary phonon modes throughout the first Brillouin zone. According to the
phonon spectrum of PG, there are three acoustic branches, two of which are longitudinal $(L)$ and transverse $(T)$ modes within the plane, and the other is an out-of-plane $(Z)$ mode. In the
vicinity of the $\Gamma$ point, the scattering of the two acoustic modes ($TA$ and $LA$) is linear, while
the scattering of the out-of-plane mode ($ZA$) is quadratic ($q_{2}$). Furthermore, based on the
phonon density of states shown in panel (b), it can be observed that the $C_{2}$ atoms with $sp^{2}$
hybridization have the highest contribution to the formation of optical phonons, where the
highest phonon density of states associated with the vibrations of $C_{2}$ atoms is approximately
at a frequency of $1600$ $cm^{-1}$. The results obtained for the calculations related to the PG
phonon spectrum are in good agreement with the results of the others  \cite{wang2016electronic}.

\subsection{Electronic Properties}

\begin{figure}[htb!]
\begin{center}
\includegraphics[width=0.90\linewidth]{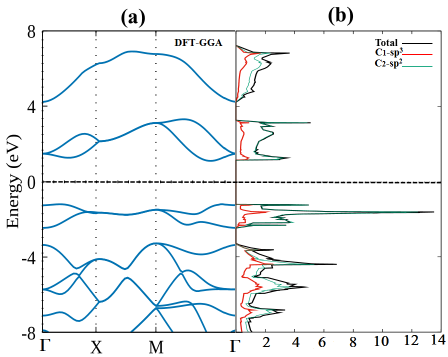}
\caption{{\textit{(Colour online) Panels (a) and (b) show the electronic band structure and density of partial
states of PG monolayer, respectively calculated with GGA-PBE functional. Fermi level is considered
at zero point.}}}
\label{FIG:03}
\end{center}
\end{figure}

To investigate the electronic properties of PG monolayer, we calculated its band structure using two approximations, LDA and GGA-PBE, at the DFT level, and the GGA-PBE approximation at the $G_{0}W_{0}$ level along high-symmetry $k-$points. Panel (a) in Fig. \ref{FIG:03} shows the band structure of PG in the GGA-PBE approximation. The Fermi energy is considered at the
zero point and the coordinates of the high-symmetry $k-$points at $\Gamma(0,0,0)$, $X(0.5,0,0)$, and $M(0.5,0.5,0)$.

\begin{figure}[htb!]
\begin{center}
\includegraphics[width=0.90\linewidth]{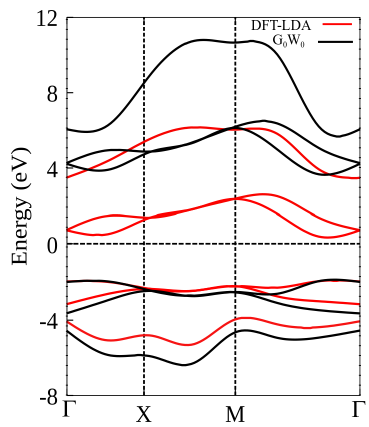}
\caption{{\textit{(Colour online) Plot shows the PG electronic band structure calculated by DFT-LDA (red
curve) and $G_{0}W_{0}$ (black curve).}}}
\label{FIG:04}
\end{center}
\end{figure}

As observed, in the DFT-GGA level, PG monolayer is a semiconductor with an
indirect band gap of approximately $2.27$ eV, with the maximum valence band occurring along
the $\Gamma-X$ path at an energy of $1.19$ eV and the minimum conduction band occurring along the
$M-\Gamma$ path at an energy of $1.08$ eV. To determine the contribution of carbon atoms with
different hybridizations in the formation of electronic density of states, we plotted the partial
density of states of the PG monolayer in panel 3(b). According to the figure, $C_{2}$ atoms with
$sp^{2}$ hybridization have the highest contribution to the formation of electronic density of
states, both in the valence and conduction bands. To obtain better and more realistic results, it
is necessary to consider the intra-structure interactions. For this purpose, we apply the $G_{0}W_{0}$
approximation to account for electron-electron interactions. As shown in Fig. \ref{FIG:04}, the use of the
$G_{0}W_{0}$ approximation leads to a wider band gap, which becomes quasi-direct with a value of
$5.3$ eV. This increase in the band gap is due to the self-energy corrections resulting from the
electron-electron interactions. As shown in Table \ref{TAB:01}, the results extracted in this study are in
good agreement with the results of the others.

\begin{table}[hbt!]
    \centering
    \caption{The values of the band gap are extracted from our calculations, which are compared with the
similar results of the others.}
    \begin{tabular}{|c|c|c|}
 \hline
 Structure  & Band gap: our results & Band gap: the others results   \\
 \hline \hline
  & DFT-LDA:
$2.25$ eV & LDA: $2.22$ eV [27]  \\ 
 Penta-graphene monolayer & DFT-LDA:
$2.32$ eV & LDA: $2.27$ eV [43]   \\
  & DFT-LDA:
$5.53$ eV & LDA: $4.48$ eV [23]  \\
 \hline
\end{tabular}
        \label{TAB:01}
\end{table}

\subsection{Optical Properties}

In calculations related to the optical properties of the selective structures, considering the
ground state properties alone is not sufficient for interpreting or predicting the experimental
results such as light propagation, absorption coefficient, loss coefficient, and the other optical parameters. Therefore, it is necessary to consider the effects of the excited states. To obtain
the optical properties of a structure, the behavior of various optical quantities as a function of
the incident photon energy needs to be studied.

\begin{figure}[htb!]
\begin{center}
\includegraphics[width=0.90\linewidth]{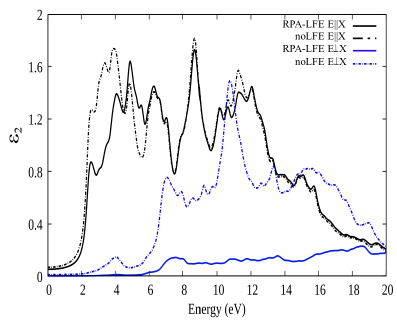}
\caption{{\textit{(Colour online) Plot shows the imaginary part of the dielectric function of the PG monolayer.
Taking into account the LFE, black solid curve illustrates the polarization of light parallel to the PG
plane. The dash-dotted black curve shows the same situation without LFE. Taking into account the
LFE, blue solid curve illustrates the polarization of light perpendicular to the PG plane. The dash-
dotted blue curve shows the same situation without LFE.}}}
\label{FIG:05}
\end{center}
\end{figure}

\begin{figure}[htb!]
\begin{center}
\includegraphics[width=0.90\linewidth]{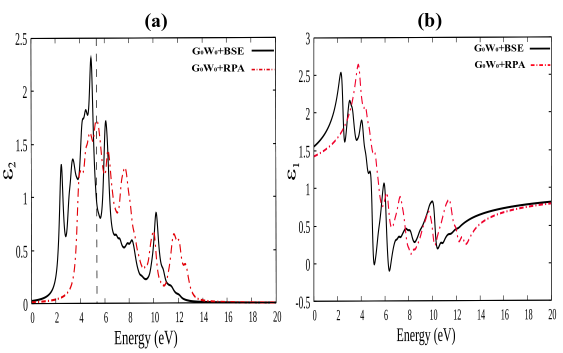}
\caption{{\textit{(Colour online) Panels (a) and (b) illustrate the imaginary and real parts of the dielectric
function of PG monolayer, respectively for the light polarization parallel to the PG plane (E||X) in the
presence (black curve) and in the absence (dash-dotted red curve) of the electron-hole interaction.
From zero energy to the vertical dotted line in the imaginary part shows the size of band gap resulting
from G0W0 approximation.}}}
\label{FIG:06}
\end{center}
\end{figure}

Calculating the dielectric function $\varepsilon_{M} = \varepsilon_{1}+i\varepsilon_{2}$ which includes both real and imaginary parts is a starting point for describing the optical
properties of the structure. Here, the dielectric function is obtained by considering the local
field effects ($LFE$), which are determined by the off-diagonal elements in the dielectric
matrix. These effects strongly modify the total response to the light polarization of the
external perturbation  \cite{ashhadi2014quasi}. Our results show that the effects of the local fields on the
calculated optical absorption spectra are negligible for parallel light polarization. It is shown
that \cite{wei2012strong} the excitonic effects by considering the local-field effects in the optical absorption
spectra of new type of hydrogenate graphene, finding that these local-field effects are highly
important. Furthermore, LFEs, which can be vital for the light polarization perpendicular to
the surface plane, are not neglected. It should be noted that the results related to the optical
properties differ significantly between the cases of light polarization parallel and
perpendicular to the plane. In the presence (absence) of the LFEs, Fig. \ref{FIG:05} illustrates the
absorption spectrum of a PG monolayer for two polarization states. The results indicate a
significant difference in the absorption spectrum due to the application of LFEs for the
polarization perpendicular to the plane compared to when these effects are not applied. When
the LFEs are applied, the absorption spectrum is shifted towards higher energies (blue shift)
and the absorption intensity is significantly reduced. On the other hand, in the case of light
polarization parallel to the PG plane, significant variations in the absorption spectrum are not
observed when we consider or neglect the LFEs. The similar results have been reported in the
other structures \cite{ashhadi2014quasi}. Therefore, here, the light polarization parallel to the plane of PG is
used. To obtain the dielectric function of PG, we first use the random phase approximation
(RPA) based on the quasi-particle energy, where only the screened electron-electron
interactions are considered ($G_{0}W_{0}$-RPA approximation). Then, we use the Bethe-Salpeter
equation approach based on the two-particle Green's function, where in addition to electron-
electron interactions, electron-hole interactions are also considered ($G_{0}W_{0}$-BSE). Fig. \ref{FIG:06}
shows the imaginary part and the real part of the dielectric function for PG monolayer in the
presence of electron-hole interaction ($G_{0}W_{0}$-BSE) and also in the absence of electron-hole
interaction ($G_{0}W_{0}$-RPA). As can be seen from panel (a) in Fig. \ref{FIG:06}, the application of electron-
hole interaction causes the absorption spectrum to shift towards lower energies (red shift)
compared to the absorption spectrum obtained from the $G_{0}W_{0}$-RPA approximation, and an important physical effect of this is the appearance of a number of bound excitons below the
bandgap resulting from the GW approximation. The optical gap of PG, which is indicated by
the first peak in the absorption spectrum and is consistent with the bound exciton, is $2.46$ eV,
and the binding energy of this exciton (defined as the difference between the quasi-particle
bandgap and the exciton energy) is $3.07$ eV. As shown in panel (b), the real part of the
dielectric function is positive above zero energy when the $G_{0}W_{0}$-RPA approximation is
applied, but with the application of electron-hole interaction, its value becomes negative
around 6 eV, indicating that no light is emitted in this energy range. The value of the real part
of the dielectric function at zero energy is defined as the static dielectric constant, which was
obtained as $1.41$ eV for the $G_{0}W_{0}$-RPA approximation and $1.52$ eV for the $G_{0}W_{0}$-BSE
approximation, indicating an increase in the dielectric constant when using the $G_{0}W_{0}$-BSE
approximation compared to the $G_{0}W_{0}$-RPA approximation, which has also been reported in
other works \cite{shahrokhi2016quasi}. In order to describe the other optical properties of the PG monolayer, the
complex refractive index has been used, which is expressed by the following equation,

\begin{eqnarray}
N(\omega) = n(\omega) + ik(\omega),
\end{eqnarray}
in which the real and imaginary parts represent the refractive index and the extinction
coefficient, respectively. These two parts are calculated using the following equations \cite{ghojavand2020ab},
\begin{eqnarray}
&&n(\omega) = \left (  \frac{\sqrt{\varepsilon _{1}^{2}\varepsilon _{2}^{2}} + \varepsilon_{1}}{2} \right )^{\frac{1}{2}} ~~ and ~~ \nonumber\\
&&k(\omega) = \left (  \frac{\sqrt{\varepsilon _{1}^{2}\varepsilon_{2}^{2}} - \varepsilon _{1}}{2} \right )^{\frac{1}{2}}.
\end{eqnarray}

\begin{figure}[htb!]
\begin{center}
\includegraphics[width=0.90\linewidth]{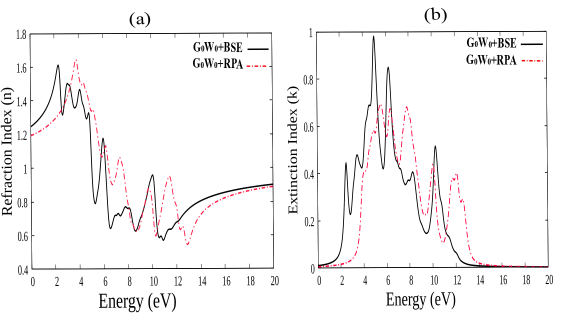}
\caption{{\textit{(Colour online) Panels (a) and (b) illustrate the refractive index n and the extinction coefficient
k of the PG monolayer, respectively in the presence (black curve) and in the absence (dash-dotted red
curve) of the electron-hole interaction.}}}
\label{FIG:07}
\end{center}
\end{figure}

\begin{figure}[htb!]
\begin{center}
\includegraphics[width=0.90\linewidth]{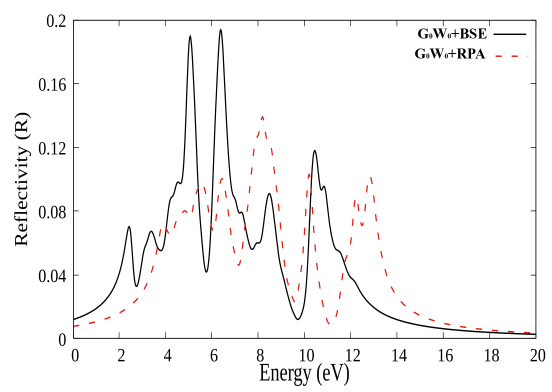}
\caption{{\textit{(Colour online) Plot shows the reflection spectra of the PG monolayer in the presence (black
curve) and in the absence (dash-dotted red curve) of the electron-hole interaction.}}}
\label{FIG:08}
\end{center}
\end{figure}

Fig. \ref{FIG:07} illustrates the calculated results for the refractive index and extinction coefficient of the
PG monolayer in the presence (absence) of the electron-hole interaction. The refractive index
at zero energy, known as the static refractive index, is introduced. According to panel (a) in
Fig. \ref{FIG:07}, this value is $1.20$ for the $G_{0}W_{0}$-RPA approximation and $1.25$ for the $G_{0}W_{0}$-BSE
approximation. Additionally, the maximum refractive index values for these two
approximations are observed at $3.81$ eV for the $G_{0}W_{0}$-RPA approximation and $2.31$ eV for
the $G_{0}W_{0}$-BSE approximation. In other studies \cite{shahrokhi2016quasi}, it has been observed that the static
refractive index calculated using $G_{0}W_{0}$-BSE is higher than the value obtained from $G_{0}W_{0}$-
RPA. Panel (b) in Fig. \ref{FIG:07} shows the extinction coefficient of the PG monolayer, which
demonstrates the electron-hole interactions cause a shift of the extinction coefficient plot
towards lower energies (red shift). The maximum extinction coefficient for the $G_{0}W_{0}$-RPA
and $G_{0}W_{0}$-BSE approximations is $0.70$ and $0.99$, respectively, which is observed at energies of $5.53$ eV and $4.98$ eV. At these energies, photons are rapidly absorbed. To proceed, the
reflectance is calculated using the following equation \cite{berdiyorov2016first111},
\begin{eqnarray}
R(\omega) = \frac{(n-1)^{2}+k^{2}}{(n+1)^{2}+k^{2}}.
\end{eqnarray}
The reflectance spectrum, which represents the intensity of incident light reflection, is shown
in Fig. \ref{FIG:08}. The maximum calculated reflectance value in the $G_{0}W_{0}$-RPA approximation is
observed at $8.71$ eV with a value of approximately $14$\%. Similarly, for the $G_{0}W_{0}$-BSE
approximation, this value is approximately $19$\% at the energy of $6.35$ eV. As evident from the
Fig. 8, the application of electron-hole interactions leads to an increase in the maximum reflectance and a shift of the reflectance spectrum towards lower energies.

\section{Conclusions and remarks}
\label{conclusion}

In conclusion, based on the presented methodology in section \ref{Methodology}, we have investigated the
electronic and optical properties of penta-graphene (PG) monolayer. The results at the DFT-
GGA level indicate that PG is a semiconductor with an indirect band gap of approximately
$2.32$ eV. The utilization of the G0W0 approximation leads to an increase in the band gap,
resulting in a quasi-direct gap of $5.35$ eV. The GW band structure calculations showed that
self-energy corrections were needed to obtain accurate quasi-particle properties for PG
monolayer. Additionally, we have employed the $G_{0}W_{0}$-RPA and $G_{0}W_{0}$-BSE approximations
to calculate the optical properties and investigate the variations in the different optical spectra
in the presence and in the absence of electron-hole interactions. The results demonstrate that
the inclusion of electron-hole interaction causes a red-shift of the absorption spectrum
towards lower energies compared to the spectrum obtained from the $G_{0}W_{0}$-RPA
approximation. Because of the enhanced overlap between electrons and holes due to the low-
dimensionality and the weak electronic screening effects in PG monolayer, optical absorption
spectra were dominated by the first bound exciton with a significant binding energy, i.e., $3.07$
eV. It is found that the PG monolayer, with a wider band gap and the enhanced excitonic
effects, holds promise as a suitable candidate for the design and fabrication of electronic and
optoelectronic components.
\newpage
\newpage
\newpage
\newpage

\newpage
\newpage

\newpage
\newpage

\newpage
\newpage

\newpage
\newpage

\newpage
\newpage

\newpage
\newpage

\section{Acknowledgements}

B.M., S.A.K. and J.M.S. acknowledges School of Physics, Damghan University. This work was supported in part by the Brazilian Agencies CAPES, CNPq, FAPESP and FAPEPI.  J.M.S acknowledges CENAPAD-SP (Centro Nacional de Alto Desenpenho em São Paulo - Universidade Estadual de Campinas - UNICAMP) provided computational support (proj842). J.M.S. acknowledges CNPq (Process No. $305053/2023-0$) for financial support.
\newpage
\bibliographystyle{elsarticle-num}
\bibliography{bibliografia.bib}

\newpage


\end{document}